\documentclass[aps,prl,twocolumn,showpacs,preprintnumbers,amsmath,amssymb]{revtex4}
\usepackage{graphicx}% Include figure files
\usepackage{dcolumn}% Align table columns on decimal point
\usepackage{bm}% bold math

\newcommand {\beq} {\begin{equation}}
\newcommand {\eeq} {\end{equation}}
\newcommand {\beqa}{\begin{eqnarray}}
\newcommand {\eeqa}{\end{eqnarray}}

\newcommand {\tr}{{\rm tr\,}}

\newcommand {\aeps}{a}

\begin{document}

%%%%%%%%%%%%%%%%%%%%%%%%%%%%%%%%%%%%%%%%%%%%%%%%%%%%%%%%%%%%%%%%%%%%
%  TITLE / AUTHOR                                                  %
%%%%%%%%%%%%%%%%%%%%%%%%%%%%%%%%%%%%%%%%%%%%%%%%%%%%%%%%%%%%%%%%%%%%

\title{The Large $N$ Reduction in Matrix Quantum Mechanics \\
--- a Bridge between BFSS and IKKT ---}

\author{Naoyuki Kawahara$^{1,2}$}
\email{kawahara@post.kek.jp}
%\cite{EmailKN}
\author{Jun Nishimura$^{1,3}$}
\email{jnishi@post.kek.jp}
%\cite{EmailJN}

%\address{
\affiliation{
$^{1}$High Energy Accelerator Research Organization (KEK), 
		Tsukuba 305-0801, Japan \\
$^{2}$Department of Physics, Kyushu University, Fukuoka 812-8581, Japan \\ 
$^{3}$Department of Particle and Nuclear Physics,
School of High Energy Accelerator Science,\\
Graduate University for Advanced Studies (SOKENDAI),
Tsukuba 305-0801, Japan
%${}^3$Department of Particle and Nuclear Physics, 
%		The Graduate University for Advanced Studies (SOKENDAI),
%			Tsukuba, Ibaraki 305-0801, Japan
}

\date{May, 2005; preprint: KEK-TH-1018, KYUSHU-HET-84,
hep-th/0505178
%\today %%new
}% It is always \today, today,

%%%%%%%%%%%%%%%%%%%%%%%%%%%%%%%%%%%%%%%%%%%%%%%%%%%%%%%%%%%%%%%%%%%%
%  ABSTRUCT    	                                                   %
%%%%%%%%%%%%%%%%%%%%%%%%%%%%%%%%%%%%%%%%%%%%%%%%%%%%%%%%%%%%%%%%%%%%

\begin{abstract}
The large $N$ reduction 
%discovered by Eguchi and Kawai 
is an equivalence between large $N$ gauge theories 
and matrix models 
%obtained by the dimensional reduction to a point.
discovered by Eguchi and Kawai in the early 80s.
In particular the continuum version of the 
quenched Eguchi-Kawai model may be useful in studying 
supersymmetric and/or chiral gauge theories nonperturbatively.
%Here we 
%advocate the usefulness of this model
%in studying the BFSS Matrix Theory,
We apply this idea to matrix quantum mechanics,
which is relevant, for instance, 
to nonperturbative studies of the BFSS Matrix Theory,
a conjectured nonperturbative definition of M-theory.
In the bosonic case we present Monte Carlo results 
confirming the equivalence directly,
and discuss a possible explanation based on the Schwinger-Dyson equations.
In the supersymmetric case we argue that 
the equivalence holds as well although some care should be taken
if the rotational symmetry is spontaneously broken.
% as speculated in the IKKT model. 
This equivalence provides an explicit relation
between the BFSS model and the IKKT model, 
which may be used to translate results in one model to the other.
\end{abstract}

%11.25.-w Theory of fundamental strings
%11.25.Sq Nonperturbative techniques; string field theory

\maketitle

%%%%%%%%%%%%%%%%%%%%%%%%%%%%%%%%%%%%%%%%%%%%%%%%%%%%%%%%%%%%%%%%%%%%
%  1. INTRODUCTION                                                 %
%%%%%%%%%%%%%%%%%%%%%%%%%%%%%%%%%%%%%%%%%%%%%%%%%%%%%%%%%%%%%%%%%%%%

\paragraph*{Introduction.---}
The large $N$ reduction \cite{EK} states 
that large $N$ gauge theories
are equivalent to matrix models obtained by
dimensionally reducing those theories to a point.
%(See Refs.\cite{review} for comprehensive reviews.)
Recent developments in the exact calculation of the low-energy superpotential
in ${\cal N}=1$ supersymmetric gauge theories \cite{DV} may also be 
regarded as
a particular application of the large $N$ reduction \cite{Kawai:2003yf}.
We apply the original idea to matrix quantum mechanics,
which is relevant, for instance, to nonperturbative studies of
the BFSS Matrix Theory \cite{BFSS} 
proposed
%by Banks, Fischler, Shenker and Susskind (BFSS) 
as a nonperturbative definition of M-theory.
The same model may also be interpreted as a low-energy effective theory
of D0-branes in type IIA superstring theory.
% \cite{witten}.
%but we discuss the model as such without specifying its interpretation.
(In the cases where fermions are absent,
interesting results
%,finiteT}
are already obtained by a direct lattice approach \cite{latticeBFSS}.
See Ref.\ \cite{Kaplan:2005ta} for a lattice formulation 
in the supersymmetric case.)

%% Here we also aim at studying supersymmetric gauge theories 
%% nonperturbatively,
%% but we take a more conventional path, which is expected to make more
%% general quantities and more general models accessible.
%% In view of the well-known difficulties in the lattice approach in the
%% presence of supersymmetry 
%% (See, however, the recent developments \cite{lattice-susy}),
%% we consider it worth while to apply a different approach albeit in the 
%% large $N$ limit.

%In realizing the Eguchi-Kawai (EK) equivalence,
 
In Ref.\ \cite{EK} it was conjectured 
that the U($N$) gauge theory on an infinite lattice
is equivalent in the large $N$ limit to a one-site model
called the Eguchi-Kawai (EK) model \cite{EK}.
This conjecture was based on the observation that
the Schwinger-Dyson equations of the two theories
coincide under the assumption that the 
U(1)$^d$ symmetry $U_{\mu} \rightarrow e^{i\alpha_{\mu}} U_{\mu}$ 
of the EK model remains unbroken.
The U(1)$^d$ symmetry, however, turned out to be spontaneously broken
in the weak coupling region \cite{BHN,Okawa:1982ic}, 
and the quenched EK (QEK) model was proposed as a remedy \cite{BHN}.
The ``quenched'' eigenvalues of $U_{\mu}$ were
%% fixed while
%% one takes the average with respect to the remaining degrees of freedom,
%% and the integration over the eigenvalues is finally performed 
%% using a uniform measure.
interpreted later as lattice momenta \cite{parisi},
and this interpretation enabled an extension to non-gauge theories.
The equivalence for the QEK model
was confirmed both perturbatively \cite{DW}
and nonperturbatively \cite{BHN,Okawa:1982nn}.
%Recently the QEK model has been used to calculate
%meson correlation functions is planar QCD \cite{Kiskis:2002gr}.

The continuum version of the QEK (cQEK) model has been proposed
in Ref.\ \cite{GK}.
%%  and it was shown by explicit
%% one-loop calculations that the gauge invariance holds in the 
%% large $N$ limit before the integration over the quenched momenta 
%% is performed.
In spite of its potential usefulness in studying
supersymmetric and/or chiral gauge theories,
the EK equivalence for the cQEK model has not been checked directly
so far.
%investigated beyond perturbation theory.
%This is one of the main issues we address in this work.

The twisted EK model \cite{TEK},
which was proposed as an alternative way to cure the problem
of the original EK model, has been interpreted recently as 
field theories on a noncommutative geometry \cite{NCYM}.
%and used for numerical simulations in this context \cite{NCsim}.
However, this model is not useful for our purpose
since the twisting procedure is applicable 
only to even space-time dimensions,
and moreover the continuum version can be defined only 
in the strict $N = \infty$ limit \cite{Gonzalez-Arroyo:1983ac}.
As another recent development, it has been found that
the breaking of the U(1)$^d$ symmetry 
can be avoided by keeping the size of the lattice
sufficiently large, and that the critical size 
in physical units becomes finite in the continuum 
limit \cite{cont-reduction}.
%% This phenomenon is referred to as the ``large $N$ continuum reduction''
%% in the literature, but it should not be confused with 
%% what we call the ``continuum version'' of the EK models.

%% for various reasons. Firstly, the twisting procedure is applicable 
%% only to even space-time dimensions.
%% Secondly, the continuum version can be
%% defined only in the strict $N = \infty$ limit \cite{Gonzalez-Arroyo:1983ac}.
%% Thirdly the spontaneous breakdown of the $U(1)^d$ symmetry, 
%% which invalidates
%% the equivalence, may occur at intermediate coupling \cite{Ishikawa-Okawa}.

%% As a no $U(1)^d$ SSB model, quenched EK (QEK) model
%% %On the other hand, 
%% %another modified EK model, 
%% %which called quenched EK (QEK) model, 
%% is proposed by 
%% Banot, Heller and Neuberger \cite{BHN}.
%% This model is investigated by Monte-Carlo simulation,
%% and obtained that the results 
%% be coincident to large $N$ lattice gauge theory, 
%% for example, by Okawa \cite{okawa}.

In this letter we consider the cQEK model
%, or the cQEK model in brief, 
for matrix quantum mechanics, and discuss its EK equivalence.
%% In fact the corresponding cQEK model looks quite similar
%% to the IKKT model \cite{IKKT}, which is proposed as a nonperturbative
%% definition of the type IIB superstring theory.
%% We consider that
%% the cQEK model is useful for Monte Carlo studies of the M-theory,
%% although in the cases where fermions are omitted (or can be omitted
%% in some circumstances), a direct lattice approach already
%% produced interesting results \cite{latticeBFSS,finiteT}.
In fact the corresponding cQEK model is analogous
to the IKKT model \cite{IKKT} proposed 
%by Ishibashi, Kawai, Kitazawa and Tsuchiya (IKKT) 
as a nonperturbative formulation of 
type IIB superstring theory.
Based on this observation, we discuss
an explicit relation between the BFSS model and the IKKT model.
%% the only difference being that the eigenvalues of 
%% one of bosonic matrices have to be quenched.
%% We speculate on a possible relation
%% between the two models, which may be used to translate results
%% in one model to the other.

%%%%%%%%%%%%%%%%%%%%%%%%%%%%%%%%%%%%%%%%%%%%%%%%%%%%%%%%%%%%%%%%%%%%
%  2. THE MODEL                                                    %
%%%%%%%%%%%%%%%%%%%%%%%%%%%%%%%%%%%%%%%%%%%%%%%%%%%%%%%%%%%%%%%%%%%%

\paragraph*{The models.---}
We study a one-dimensional U($N$) gauge theory,
or matrix quantum mechanics (QM),
with adjoint scalars $X_i(t)$  $(i=1,\cdots,d-1)$, which are 
$N\times N$ Hermitian matrices. The action is given by
\begin{equation}
S_{\rm mQM} = 
\frac{1}{g^2} \int 
% _{-\infty} ^{\infty} 
d t \, 
\tr \left\{ 
\frac{1}{2} (D X_i(t))^2 - 
\frac{1}{4} [X_i(t),X_j(t)]^2
\right\} \ ,
 \label{cQM}
\end{equation}
where $D X_i = \frac{\partial}{\partial t} 
X_i - i \, [A,X_i]$ represents the covariant derivative
with $A$ being the 1d gauge field.
%The integration range $\tau$ is sent to infinity.
The model can be obtained formally by the dimensional reduction
of $d$-dimensional Yang-Mills theory to one dimension.
%:
%\begin{eqnarray*}
%A_{\tau} & \rightarrow & 
%gA_{\tau}g^{\dagger}-ig\partial_{\tau}g^{\dagger} \\
%X_i & \rightarrow & gX_ig^{\dagger}
%\end{eqnarray*}
%where $g(\tau) \in SU(N)$.
In what follows we fix the gauge to $A(t)=0$.
%\cite{endnote1}.
This model is UV finite and hence
it does not require any UV regularization.
As a result, the parameter $g$ in (\ref{cQM})
can always be scaled out by
an appropriate rescaling of the matrices and 
the time coordinate $t$.
We take $g = \frac{1}{\sqrt{N}}$, which turns out to be
convenient when we discuss the large $N$ limit.

%\textit{}
Applying the quenching prescription \cite{parisi,GK}
to the matrix QM (\ref{cQM}),
%% \begin{eqnarray}
%% X_i(t)  & \mapsto & e^{i \Omega t}  \,
%% X_i \, e^{-i \Omega t}  \ ,
%% %% \frac{\partial}{\partial t} X_i(t)
%% %%  & \mapsto & i \, [\Omega,X_i] \nonumber \\
%% %% %  = \omega_{a,\mu} \delta_{a,b}
%% %% \int d t & \mapsto & \frac{2\pi}{\Lambda} \equiv \epsilon \ ,
%% \label{QMP}
%% \end{eqnarray}
we obtain the corresponding cQEK model
\begin{equation}
S_{{\rm  cQEK},\Omega} = - N \, \epsilon \, \tr \left( 
\frac{1}{2} \,  [\Omega,X_i]^2
+  \frac{1}{4} \,  [X_i,X_j]^2 \right) \ ,
\label{cQEK}
\end{equation}
where $X_i$ ($i=1 , \cdots , d-1$) are $N \times N$ 
traceless Hermitian matrices and $\Omega
\equiv \mbox{diag}(\omega_1 , \cdots , \omega_N)$
represents frequency variables to be integrated over 
the region $[-\frac{\Lambda}{2} , \frac{\Lambda}{2}]$,
where $\Lambda \equiv \frac{2\pi}{\epsilon}$.
%$| \omega_i | \le \frac{\Lambda}{2}$ with
%the ultraviolet cut-off $\Lambda=\frac{2\pi}{\epsilon}$.
%(its inverse is lattice spacing $\epsilon$)
(More precisely, the difference $\omega_i - \omega_j$ 
corresponds to the frequency.)

As U($N$) invariant operators in the matrix 
QM (\ref{cQM}), let us consider
%\begin{equation}
$
\mathcal{O} =  \frac{1}{N} \tr X_{i_1}(t_1) \cdots X_{i_n}(t_n)  
$
%\label{obs-QM}
%\end{equation}
and the corresponding operator
%\begin{equation}
$
\mathcal{\tilde{O}} =  \frac{1}{N} \tr \tilde{X}_{i_1}(t_1) \cdots 
\tilde{X}_{i_n}(t_n)
$
%\label{obs-cQEK}
%\end{equation}
in the cQEK model (\ref{cQEK}),
where 
$\tilde{X}_{i}(t) \equiv e^{i\Omega t} \, {X}_{i} \, e^{-i\Omega t}$.
%
%As a fundamental quantity, let us consider 
%$\langle \frac{1}{N} \tr (X_i(t))^2 \rangle$.
%% \begin{equation}
%% {\cal O}=
%% \frac{1}{\tau} \int dt _{-\tau/2}^{\tau/2}
%% \tr \{X_i(t)\}^2 \ .
%% \end{equation}
Then, according to the EK equivalence,
\begin{equation}
\lim_{N \rightarrow \infty} \langle\mathcal{O}\rangle_{\rm mQM} = 
\lim_{\Lambda \rightarrow \infty}
\lim_{N \rightarrow \infty} 
\int d \Omega \,
\langle\mathcal{\tilde{O}}\rangle_{{\rm cQEK},\Omega} \ ,
\label{QM=QEK}
\end{equation}
where $\langle \ \cdot  \ \rangle_{\rm mQM}$ 
and $\langle \ \cdot \ \rangle_{{\rm cQEK},\Omega}$ 
represent the expectation values
with respect to the actions (\ref{cQM}) and (\ref{cQEK}),
respectively, and 
%the integration over $\Omega$ is defined by
%\begin{equation}
$
\int d \Omega \equiv
\prod_{i=1}^{N} 
\int^{\frac{\Lambda}{2}}_{-\frac{\Lambda}{2}}
\frac{d\omega_i}{\Lambda}
% \ .
$.
%\end{equation}
The order of the two limits on the r.h.s.\ of (\ref{QM=QEK}) cannot
be inverted.

%%%%%%%%%%%%%%%%%%%%%%%%%%%%%%%%%%%%%%%%%%%%%%%%%%%%%%%%%%%%%%%%%%%%
%  5. MONTE CARLO SIMULATION                                       %
%%%%%%%%%%%%%%%%%%%%%%%%%%%%%%%%%%%%%%%%%%%%%%%%%%%%%%%%%%%%%%%%%%%%

\paragraph*{Monte Carlo simulation.---}

We test the equivalence (\ref{QM=QEK}) directly
by Monte Carlo simulation.
In order to simulate the 1d model (\ref{cQM}),
we discretize the time ``$t$'',
and obtain the action
%we need to discretize the time ``$t$''.
%This can be done straightforwardly in the present bosonic model,
%and we obtain the action
\begin{eqnarray}
S_{\rm lat} & = & N \aeps
\left\{
\sum_{n=1}^{T-1} 
\frac{1}{2} \, \tr \! \left(\frac{X_i(n+1)-X_i(n)}{\aeps}\right)^2  
\right.
\nonumber \\
& {} & 
-\sum_{n=1}^{T} 
\left.
 \frac{1}{4} \, \tr [X_i(n),X_j(n)]^2
\right\}   \ ,
\label{latticeQM}
\end{eqnarray}
where $\aeps$ represents the lattice spacing.
The $N\times N$ Hermitian matrices $X_i(n)$ ($n=1,\cdots , T$) sit
on each site,
%of the 1d lattice
and we do {\em not} impose
periodic boundary conditions.
%; {\em i.e.}, 
%$X_i(1)$ and $X_i(T)$ are treated as independent variables.
%assume the free boundary condition. 
In order to retrieve the original model (\ref{cQM}), 
we have to take the $\aeps \rightarrow 0$
and $\tau  \equiv \aeps \, T \rightarrow \infty$
limits, where the order of the limits should not matter.
The system is invariant under 
$X_i(n) \mapsto X_i(n) + \beta_i  \, {\bf 1}$,
and we fix the corresponding zero mode
% which actually decouples from the other
%d.o.f.\ 
by imposing the condition
$\sum_{n=1}^{T} \tr X_i(n) = 0$.

%% The discretization of ``t'' in the supersymmetric case is not that 
%% straightforward due to the appearance of the fermion doublers.
%% However, we may add a Wilson term such as
%% \begin{eqnarray}
%% \epsilon w \sum_{n=2}^{T-1}
%% \frac{1}{2} {\cal A}_{\alpha\beta} \Psi_\alpha
%% \left(\frac{\Psi_\beta(n+1)-2 \Psi_\beta(n)+\Psi_\beta(n-1)}
%% {\epsilon^2}\right)^2   \ , 
%% \label{Wilson-term}
%% \end{eqnarray}
%% which removes the unwanted doublers. 
%% The SO($d$) symmetry, which is broken on the lattice due to
%% the Wilson term, is expected to be restored
%% in the continuum limit since the term (\ref{Wilson-term})
%% vanishes in the classical continuum limit, and there is no 
%% room for anomaly due to the UV finiteness.
%% (Supersymmetry is also expected to be restored 
%% in the continuum limit.)

Monte Carlo simulation of the models
(\ref{latticeQM}) and (\ref{cQEK}) can be performed by
the heat bath algorithm \cite{HNT}.
% in much the same way
%as in the bosonic IKKT model \cite{HNT}.
%
%% The integration over $\Omega$ required for the cQEK model is 
%% performed by using random variables distributed
%% uniformly within the region $[-\Lambda/2 , \Lambda/2]$.
%% %and repeating the simulation of (\ref{cQEK}) for each $\Omega$.
%% Thus in order to obtain results for the cQEK model,
%% we first take an average the ensemble $\{X_i\}$ with 
%% the action (\ref{cQEK}) 
%% for each $\Omega$, and then perform the averaging over $\Omega$
For the cQEK model,
we first take an average over the ensemble $\{X_i\}$ with 
the action (\ref{cQEK}) for each $\Omega$, 
and then perform the averaging over $\Omega$
by using random numbers distributed
uniformly within the region $[-\frac{\Lambda}{2} , \frac{\Lambda}{2}]$.
Correspondingly there are two types of statistical errors.
%% The one coming from the ensemble average over $\{X_i\}$ 
%% %can be estimated by the standard jackknife method.
%% and the other one coming from the $\Omega$-integration.
%can be estimated by the standard formula for the mean square error.
We estimate the total error by
taking an average over the error of the first type and adding the
one of the second type.

\begin{figure}[htb]
\begin{center}
\includegraphics[height=5cm]{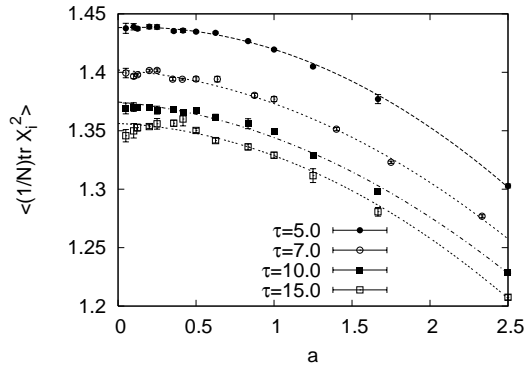}
%{bfss_continuum.eps}
\end{center}
\caption{The observable $\langle \mathcal{O}_1 \rangle_{\rm lat}$ 
for the lattice model (\ref{latticeQM}) with $N=16$
is plotted against $\aeps$
for $\tau=$ 5.0, 7.0, 10.0, 15.0.
The lines represent fits to
$\langle \mathcal{O}_1 \rangle_{\rm lat} =
C_1 + C_2 \, \aeps + C_3 \, \aeps^2$.}
\label{bfss_continuum}
\end{figure}

We focus on the $d=4$ case in what follows.
Let us first consider the observable 
$\mathcal{O}_1 = \frac{1}{NT} \sum^{T}_{n=1} 
\tr (X_{i}(n)^2) $
for the lattice model (\ref{latticeQM}).
In Fig.\ \ref{bfss_continuum} we plot the results for 
$\langle \mathcal{O}_1 \rangle_{\rm lat}$ against $\aeps$
for $\tau=$ 5.0, 7.0, 10.0, 15.0 at $N=16$.
We fit the results to $\langle \mathcal{O}_1 \rangle_{\rm lat} =
C_1 + C_2 \, \aeps + C_3 \, \aeps^2$,
from which we obtain the continuum limit.

%%%%% BFSS_continuum. %%%%%
%%%%%%%%%%%%%%%%%%

\begin{figure}[htb]
\begin{center}
\includegraphics[height=5cm]{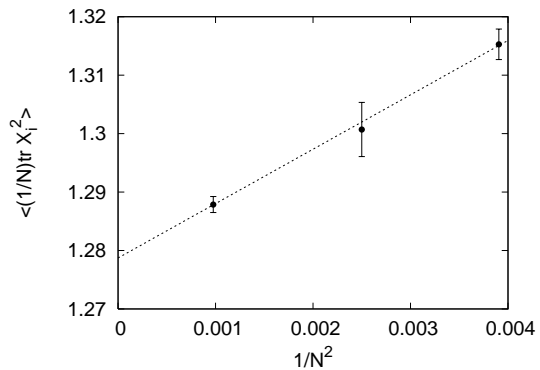}
%{bfss_continuum.eps}
\end{center}
\caption{The observable
%$\langle \mathcal{O}_1 \rangle_{\rm mQM}$ 
%$\lim_{\tau \rightarrow \infty}\lim_{\epsilon \rightarrow 0 }
$\langle \mathcal{O}_1 \rangle_{\rm lat}$ 
%for the matrix QM (\ref{cQM})
for the lattice model (\ref{latticeQM})
after taking the $\aeps \rightarrow 0$ and $\tau \rightarrow \infty$
limits are plotted against $\frac{1}{N^2}$.
The straight line represents a fit to $C_6 + C_7 \, N^{-2}$.
}
\label{bfss_largeN}
\end{figure}
%

%At FIG. \ref{bfss_thermodynamic} 
%In Fig.\ \ref{bfss_thermodynamic} we plot the results
The results obtained in the continuum limit for various $\tau$
%for $\langle \mathcal{O}_1 \rangle_{\rm lat}$ 
%in the continuum limit at $N=16$ for various $\tau$
%against $1/\tau$.
can be fitted to
$ \lim_{\aeps \rightarrow 0 }\langle \mathcal{O}_1 \rangle_{\rm lat} = 
C_4 + C_5 \, \tau^{-1}$, 
from which we obtain the $\tau \rightarrow \infty$
limit. 
We redo this analysis for $N=20$, 32.
% and plot the results against $\frac{1}{N^2}$ 
The results can be fitted to $C_6 + C_7 \, N^{-2}$
as one can see from Fig.\ \ref{bfss_largeN}.
This large $N$ behavior is analogous to the one
observed in the bosonic IKKT model \cite{HNT}.
Thus we can obtain the final result for the 
matrix QM  (\ref{cQM}).
%large $N$ limit.

%%%%%%%%%%%%%%%%%%%%%%%
%% \begin{figure}[htb]
%% \begin{center}
%% \includegraphics[height=5cm]{bfss_thermodynamic_16_a2.eps}
%% \end{center}
%% \caption{The results in the continuum limit
%% $\lim_{\epsilon \rightarrow 0} \langle \mathcal{O}_1 \rangle_{\rm lat}$ 
%% are plotted against $\tau$ at $N=16$.}
%% \label{bfss_thermodynamic}
%% \end{figure}
%%%%%%%%%%%%

\begin{figure}[htb]
\begin{center}
\includegraphics[height=5cm]{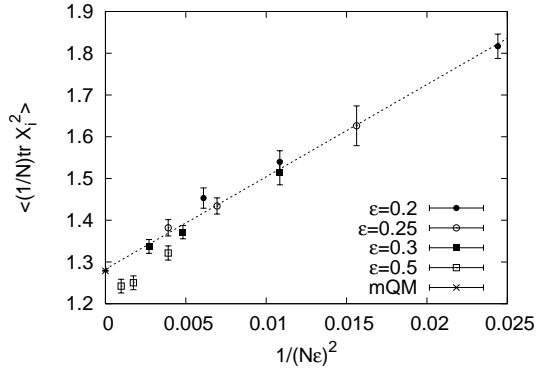}
%{qek_largeN.eps}
\end{center}
\caption{The observable
$\langle \mathcal{\tilde{O}}_1 \rangle_{\rm cQEK}$ 
for the cQEK model (\ref{cQEK})
is plotted against $\frac{1}{(N\epsilon)^2}$
for $\epsilon=$ 0.2, 0.25, 0.3, 0.5.
The straight line represents a fit to 
$\langle \mathcal{\tilde{O}}_1 \rangle_{\rm cQEK} = 
C_8 + C_{10} \, (N\epsilon)^{-2}$. The cross on the vertical axis
(the error bar is almost invisible)
represents the final result for the matrix QM (\ref{cQM}).
}
\label{qek_largeN}
\end{figure}

Let us move on to the cQEK model (\ref{cQEK}).
The corresponding observable is
$\mathcal{\tilde{O}}_1 = \frac{1}{N} 
\tr (X_{i}^2)$.
In Fig.\ \ref{qek_largeN} we plot the results
$\langle \mathcal{\tilde{O}}_1 \rangle_{\rm cQEK}$ 
obtained for $N=32$, 48, 64 with $\epsilon=$ 0.2, 0.25, 0.3, 0.5.
For fixed $\epsilon$ the large $N$ behavior
is given by $C_8 + C_9 \, N^{-2}$.
The coefficient $C_9$, however, diverges as $\frac{1}{\epsilon^2}$.
This can be understood if we recall
that the density of $\omega_i$ in
the region $[-\frac{\Lambda}{2} , \frac{\Lambda}{2}]$
is $\rho = \frac{N}{\Lambda}=\frac{N\epsilon}{2\pi}$.
Since finite $N$ effects at small $\epsilon$
come mainly from the finiteness of $\rho$, any power of $N$ should
be associated with the same power of $\epsilon$.
This motivated us to plot the results 
against $\frac{1}{(N\epsilon)^2}$
in Fig.\ \ref{qek_largeN}, where
the results indeed lie on a single straight line
for $\epsilon \le 0.3$.
Fitting all the data for $\epsilon \le 0.3$,
we obtain the final result for the cQEK model (\ref{cQEK}).

%%%%% qek_largeN. %%%%%

%%%%%%%%%%%%%%%%%%

We repeat this analysis for other observables
%% \begin{eqnarray}
%% \mathcal{O}_2 &=& \frac{1}{NT} \sum^{T}_{n=1} 
%% \tr (X_{i}(n)^2 X_{j}(n)^2) 
%% \nonumber \\
%% \mathcal{O}_3 &=& \frac{1}{NT} \sum^{T}_{n=1} 
%% \tr (X_{i}(n)X_{j}(n)X_{i}(n)X_{j}(n))  \ ,
%% %\nonumber \\
%% %\mathcal{O}_4 &=& \frac{1}{NT} \sum^{T}_{n=1} 
%% %Tr (X_{i}(n+a)X_{i}(n))
%% \label{O23}
%% \end{eqnarray}
%% in the lattice model (\ref{latticeQM}) and the corresponding 
%% ones in the cQEK model (\ref{cQEK}) 
and summarize the results
in Table \ref{result}, where we find good agreement
%between the two models 
within error bars.
We consider this as a compelling evidence for the EK equivalence
(\ref{QM=QEK}). 
%between the two models.

%%%%%%%%%%%%%%%%%%%%%%%%%%%%%%%%%%%%%%%%%%%%%%%%%%%%%%%%%%%%%%%%%%%%
%  4.SCHWINGER-DYSON EQUATIONS                                     %
%%%%%%%%%%%%%%%%%%%%%%%%%%%%%%%%%%%%%%%%%%%%%%%%%%%%%%%%%%%%%%%%%%%%

\paragraph*{Schwinger-Dyson equations.---}
Let us discuss a possible explanation for
the equivalence (\ref{QM=QEK}) based
on the Schwinger-Dyson (SD) equations \cite{EK,GK,Makeenko:2002uj}.
%, 
%in the other words, the EK equivalence in terms of
%analytic calculation.

In the matrix QM (\ref{cQM}) we consider
\begin{equation}
%I_{QM} \equiv 
\langle \tr 
( \lambda^a  X_{i_1}(t_1) \cdots X_{i_n}(t_n) ) 
\rangle _{\rm mQM} \ , \nonumber
\end{equation}
where $\lambda^a$ represents a generator of the U($N$) group.
Changing the integration variables 
$X_{i}(t) \mapsto X_{i}(t) + \lambda^a \delta_{i , i_0}
\delta(t-t_0)$,
summing over $a$, using the identity
$\sum_{a}(\lambda^a)_{ij} (\lambda^a)_{kl} = \delta_{il} \delta_{jk}$,
and assuming the large $N$ factorization property
$\langle {\cal O}{\cal O}' \rangle_{\rm mQM} \simeq
\langle {\cal O} \rangle_{\rm mQM} \langle {\cal O}' \rangle_{\rm mQM} $,
%$\langle \frac{1}{N}\tr A \frac{1}{N}\tr B \rangle =
%\langle \frac{1}{N} \tr A \rangle \langle \frac{1}{N}\tr B \rangle
%+ {\rm O} (\frac{1}{N^2})$,
we obtain a closed set of equations for the observables 
$\langle {\cal O} \rangle_{\rm mQM}$, which is the SD equations.

In the cQEK model we consider
\begin{equation}
%I_{\rm cQEK} \equiv 
\int d \Omega \, 
\langle \tr( e^{i\Omega t_0} 
\lambda^a  e^{-i\Omega t_0} \nonumber \\
\tilde{X}_{i_1}(t_1) 
\cdots \tilde{X}_{i_n}(t_n) ) \rangle _{{\rm cQEK},\Omega} \ .
\end{equation}
Changing the integration variables
$X_{i} \mapsto X_{i} + \frac{1}{\epsilon} \lambda^a \delta_{i , i_0}$, 
we proceed similarly to the matrix QM case
%except that here we have to assume 
assuming in addition that the $\Omega$-integration factorizes as well.
%the large $N$ factorization property for the $\Omega$-integration
%$\int d\Omega (AB)  = \int d\Omega A \cdot \int d\Omega B + {\rm O}
%(\frac{1}{N^2})$ as well.
In this way we obtain a set of equations, which are the same as the
SD equations for the matrix QM
with the identification (\ref{QM=QEK})
except for the terms 
($s = 1, \cdots , n$)
\begin{eqnarray}
\frac{1}{\epsilon} \delta_{i_0 , i_s} \!\! \int \!\! d \Omega
 \langle \tr ( e^{i\Omega (t_s -t_0)} \tilde{X}_{i_1}(t_1) 
\cdots \tilde{X}_{i_{s-1}}(t_{s-1}) ) \rangle 
_{{\rm cQEK},\Omega} 
\nonumber \\
\times \!\! \int \!\! d \Omega
\langle \tr (\tilde{X}_{i_{s+1}}(t_{s+1}) 
\cdots \tilde{X}_{i_n}(t_{n}) e^{-i\Omega (t_s -t_0)} ) \rangle 
_{{\rm cQEK},\Omega} \nonumber \ ,
%\label{source}
\end{eqnarray}
whose counterpart exists in the matrix QM
only when $t_s = t_0$.
If we neglect the fact that the integration region
of $\Omega$ is bounded by $\Lambda$, 
the system is invariant under the shift
$\Omega \mapsto \Omega + \alpha \, {\bf 1}$ 
except for the factors
$e^{\pm i\Omega (t_s -t_0)}$,
which make the two $\Omega$-integrations vanish separately
for $t_s \neq t_0$.
% since the integrand gives a phase factor
% $e^{\pm i\alpha(t_s - t_0)}$ under the shift of $\Omega$.
%% Since the model we are studying is UV finite,
%% we may expect that the boundary of the integration range
%% should not be important.
%% Thus we find that the Schwinger-Dyson equations for the two theories
%% coincide with each other.

Since the model we are studying is UV finite,
we may expect that the integrand for the $\Omega$-integration
is determined by the $\omega_i$'s which are clustered in a finite
region.
Then the integration over the position of the cluster 
%within the range $[-\frac{\pi}{\epsilon} , \frac{\pi}{\epsilon}]$
gives a factor of $f(t_s - t_0)=
\int _{-\Lambda/2}^{\Lambda/2} \frac{d\omega}{\Lambda}
e^{i\omega (t_s-t_0)} $,
% = \frac{\sin (\Lambda (t_s - t_0) /2)}{\Lambda t /2}$,
and we obtain the coefficient
$\frac{1}{\epsilon} f(t_s-t_0)^2\rightarrow \delta(t_s-t_0)$,
which agrees with the corresponding one 
in the matrix QM (\ref{cQM}) in the $\epsilon \rightarrow 0$ limit.
%Thus we find that the Schwinger-Dyson equations for the two theories
%coincide with each other.
Assuming that the SD equations have a unique solution,
we obtain the equality (\ref{QM=QEK}).

%those $\omega_i$ which are well within the integration region
%$[ - \frac{\pi}{\epsilon} , \frac{\pi}{\epsilon} ]$
%give the dominant contribution.

%%%%% result. %%%%%
\begin{table}
\begin{center}
\begin{tabular}{ccc} 
\hline \hline
{~~~observable~~~} & {~~~matrix QM~~~} & {~~~cQEK~~~} \\ 
\hline 
{$\langle \frac{1}{N} \tr X_i^2 \rangle$} & {1.279(1)} & {1.28(1)} \\
% \hline
{$\langle \frac{1}{N} \tr X_i^2X_j^2 \rangle$} 
& {2.15(1)} & {2.11(5)} \\ 
%\hline 
{$\langle \frac{1}{N} \tr (X_iX_j)^2 \rangle$} 
& {1.23(2)} & {1.27(4)} \\ \hline \hline
\end{tabular}
\end{center}
\caption{Comparison of the Monte Carlo results obtained
for the matrix QM (\ref{cQM})
and the cQEK model (\ref{cQEK}).}
\label{result}
\end{table}

%%%%%%%%%%%%%%%%%%%%%%%%%%%%%%%%%%%%%%%%%%%%%%%%%%%%%%%%%%%%%%%%%%%%
%  6. IKKT vs BFSS                                                 %
%%%%%%%%%%%%%%%%%%%%%%%%%%%%%%%%%%%%%%%%%%%%%%%%%%%%%%%%%%%%%%%%%%%%

\paragraph*{Relation to the IKKT model.---}
The cQEK model (\ref{cQEK}) for the matrix QM has the action of
the same form as the bosonic IKKT model \cite{Krauth:1998yu}
%\begin{eqnarray}
$\, S_{\rm bIKKT} = - \frac{N}{4} \tr [A_{\mu},A_{\nu}]^2 \, $
%  \ ,
%\label{IKKT}
%\end{eqnarray}
if we identify $A_i \equiv \epsilon ^{1/4} X_i$ 
($i \le d-1$) and $A_{d} \equiv \epsilon ^{1/4} \Omega$.
The only difference is that $A_{d}$ has to be quenched.
In the bosonic IKKT model it is known that 
$\langle\frac{1}{N}\tr (A_\mu)^2 \rangle_{\rm bIKKT}$  is of O(1)
\cite{HNT}, and the extent of the eigenvalue
distribution is of O(1) in all directions.
The quenching stretches the eigenvalue distribution of 
$A_{d}$ to have an extent of O($\epsilon^{-3/4}$),
and as a result the eigenvalue distribution of $A_i$
($i \le d-1$) shrinks to O($\epsilon^{1/4}$) 
according to our results.
This implies that, in spite of their formal similarity,
the bosonic BFSS model 
and the bosonic IKKT model
probe totally different regions of the configuration space.

How about the supersymmetric case? 
In $d=4$, since the situation is qualitatively the same
(no spontaneous breaking of the rotational symmetry and
$\langle\frac{1}{N}\tr (A_\mu)^2 \rangle \sim \mbox{O}(1)$ \cite{4dsusy}),
all the statements for the bosonic case should equally apply.

In $d=10$ (i.e., the IKKT model), there are certain evidences
that the eigenvalue distribution of $A_\mu$
collapses dynamically to a four-dimensional hypersurface \cite{SSB}.
Let us assume further that the eigenvalue distribution
extends to infinity in four directions at large $N$,
and that the distribution is uniform in these directions,
as expected if the IKKT model really describes our 4d space-time.
When we quench the eigenvalues of $A_{10}$, 
the 4d hypersurface will arrange itself to have the required
extent in the 10th direction.
In this situation 
%the 4d hypersurface is not really forced 
a vector in the 10th direction may 
have non-zero components in directions orthogonal to the 4d hypersurface,
which implies that the system is not invariant under 
a shift in the 10th direction.
We therefore suspect that the EK equivalence 
between the BFSS model and the corresponding cQEK model
does not hold {\em as it stands}.

We may, however, add a ``mass term'' 
$\frac{1}{2} N \, m^2 \int dt  \, \tr  X_{i}^2$
to the BFSS model, and consider the corresponding cQEK model,
in which the mass term forces
the 10th direction to be included in the 4d hypersurface.
This guarantees the translational invariance in the 10th direction
except at the boundary, which should be fine again due to the UV finiteness.
Therefore the EK equivalence in this case 
is expected to hold at any finite $m$,
and it should hold, too, even if we send $m$ to zero eventually.

We speculate that actually quenching is not needed
since the 10th direction, in which we do not have
the mass term, will extend
to infinity as $N\rightarrow \infty$ anyway.
This implies that the IKKT model with a small mass term
in 9 directions, which should be removed after taking the large $N$ limit,
is equivalent (in the sense of Eguchi-Kawai) to the BFSS model.
% \cite{endnote}.
In particular if the IKKT model has 4 extended directions,
the BFSS model should have 3 out of 9 transverse directions
extended, thus breaking the SO(9) symmetry down to SO(3) spontaneously.
This equivalence
% between the BFSS model and the IKKT model
%we conjecture 
is somewhat reminiscent of
the T-duality, which relates the BFSS model on a circle
to the IKKT model with Taylor's
compactifying condition \cite{Taylor:1996ik}.
Note, however, that our conjecture
is different in that the two models can be {\em both}
considered in the decompactified limit.

%%%%%%%%%%%%%%%%%%%%%%%%%%%%%%%%%%%%%%%%%%%%%%%%%%%%%%%%%%%%%%%%%%%%
%  7. SUMMARY                                                      %
%%%%%%%%%%%%%%%%%%%%%%%%%%%%%%%%%%%%%%%%%%%%%%%%%%%%%%%%%%%%%%%%%%%%

\paragraph*{Concluding remarks.---}
In this letter we have provided the first direct confirmation
of the EK equivalence for the cQEK model
in the case of matrix QM or the 1d gauge theory.
Whether it holds also in higher dimensions is an interesting 
open question.
This is relevant to nonperturbative studies of Matrix String Theory
\cite{MST},
and it is also important in view of the difficulties in formulating
supersymmetric and/or chiral gauge theories on the lattice.
(See Refs.\ \cite{lattice-susy,Kaplan:2005ta} for recent developments.)
%in ``lattice supersymmetry''.)
As is clear from the discussion on the SD equations,
the question is highly nontrivial in the presence of 
UV divergences. In Ref.\ \cite{GK} it is speculated that the
equivalence holds if the UV divergence is at most logarithmic.
%We hope to report on progress in that direction in the near future.
A natural starting point in this direction
would be to study the cQEK model for the
2d pure Yang-Mills theory, and see whether the known exact results 
%\cite{Gross:1980he}
can be reproduced.
% \cite{work-in-progress}.

%%%%%%%%%%%%%%%%%%%%%%%%%%%%%%%%%%%%%%%%%%%%%%%%%%%%%%%%%%%%%%%%%%%%
%  ACKNOWLEDGEMENTS                                                %
%%%%%%%%%%%%%%%%%%%%%%%%%%%%%%%%%%%%%%%%%%%%%%%%%%%%%%%%%%%%%%%%%%%%

\paragraph*{Acknowledgments.---}
The authors would like to thank T.\ Azuma, S.\ Iso, H.\ Kawai, Y.\ Kitazawa, 
K.\ Nagao, K.\ Ohta, F.\ Sugino, A.\ Tsuchiya and K.\ Yoshida
for valuable discussions.
The work of J.N.\ is supported in part by Grant-in-Aid for 
Scientific Research (No.\ 14740163)
from the Ministry of Education, Culture, Sports, Science and Technology. 

%%%%%%%%%%%%%%%%%%%%%%%%%%%%%%%%%%%%%%%%%%%%%%%%%%%%%%%%%%%%%%%%%%%%
%  REFFERENCE                                                      %
%%%%%%%%%%%%%%%%%%%%%%%%%%%%%%%%%%%%%%%%%%%%%%%%%%%%%%%%%%%%%%%%%%%%

%\end{references}

\end{document}